\documentclass[a4j,11pt,onecolumn,oneside,notitlepage,final]{article}

\usepackage{amsmath}
\usepackage{amsfonts}
\usepackage{amssymb}
\usepackage{latexsym}
\usepackage{color}
\usepackage{graphicx}
\usepackage{cite}
\usepackage{subfigure}
\usepackage{ascmac}

\def\be{\begin{equation}}
\def\ee{\end{equation}}
\def\ben{\begin{eqnarray}}
\def\een{\end{eqnarray}}

\begin{document}
	\begin{center}
		\vskip .5in
		
		{\Large \bf
			A novel formulation of the PBH mass function
		}
		\vskip .45in
{
Teruaki Suyama$^{a}$,
Shuichiro Yokoyama$^{b,c}$,
}

{\em
$^a$
   Department of Physics, Graduate School of Science, The University of Tokyo, Tokyo 113-0033, Japan
}\\
{\em
$^b$
   Kobayashi Maskawa Institute, Nagoya University, Aichi 464-8602, Japan
}\\
{\em
$^c$
   Kavli IPMU (WPI), UTIAS, The University of Tokyo, Kashiwa, Chiba 277-8583, Japan
}\\

\end{center}
	
\abstract{
Computations of the primordial black hole (PBH) mass function discussed in the literature have conceptual issues.
They stem from that the mass function is a differential quantity and the standard criterion of the PBH formation
from the seed primordial fluctuations cannot be directly applied to the computation of the differential quantities.
We propose a new criterion of the PBH formation which is an addition of one extra condition
to the existing one.
By doing this, we derive a formal expression of the PBH mass function without introducing any ambiguous
interpretations which exist in the previous studies.
Once the underlying primordial fluctuations are specified, 
the PBH mass function can be in principle determined by the new formula.
As a demonstration of our formulation,
we compute the PBH mass function analytically for the case where the perturbations are Gaussian
and the space is one dimension.
}

\section{Introduction}

There are multiple reasons why studies of primordial black holes (PBHs) hold a unique position in cosmology.
Firstly, PBHs provide one of a few probes for extremely small-scale primordial density fluctuations
not accessible by measurements of the Cosmic Microwave Background and Large-Scale Structure (e.g. \cite{Carr:2005zd}).
Secondly, PBHs (in some mass range) can comprise all dark matter (e.g. \cite{Carr:2016drx}).
Thirdly, PBHs may explain all of or some of the binary BHs observed by LIGO/Virgo (e.g. \cite{Sasaki:2018dmp}).

Important observables characterizing PBHs are their abundance and mass function.
So far, various observations and experiments have provided limits on the PBH abundance as well as
the PBH mass function for a wide range of the PBH masses \cite{Carr:2017jsz}.
Within the context of probing the small-scale primordial fluctuations,
given that the PBH abundance and the mass function are determined by the primordial fluctuations,
observational limits on PBHs can be then translated into the upper limit on the amplitudes 
of the primordial density fluctuations.
More precisely, non-detections of PBHs over some mass range are translated into the upper limit
on the primordial power spectrum over the range of length scales corresponding to the PBH mass range.
Once the constraints on the small-scale fluctuations are obtained, they can be used to constrain 
certain class of inflation models.
In order for the flow of this argument to work, it is necessary to know a priori how the PBH abundance and the
mass function are determined for a given primordial density fluctuations.

Computations of the PBH abundance and the mass function have been a long-term issue and
addressed by many authors in the literature
\cite{Carr:1975qj,Kim:1996hr, Yokoyama:1998xd, Green:2004wb, Lyth:2005ze, Zaballa:2006kh, Bugaev:2008gw, Carr:2009jm, Young:2014ana, Kuhnel:2015vtw, Inomata:2017okj, Bernal:2017vvn, Bellomo:2017zsr, Yoo:2018esr, Germani:2018jgr,  Byrnes:2018clq, Byrnes:2018txb, Kawasaki:2019mbl, Young:2019yug, Wang:2019kaf, Young:2019osy, MoradinezhadDizgah:2019wjf,Kalaja:2019uju}.
So far, several methods to compute those quantities have been proposed based on either the 
Press-Schechter formalism or the peak-theory formalism. 
Nevertheless, as we will explain in detail in section \ref{previous}, 
previous methods have issues at the conceptual level.
Essential point common to these issues is that the PBH mass function is a differential quantity
and criterion of the PBH formation
from the seed random fluctuations has not been formulated manifestly in a way directly applicable
to the computation of the differential quantities.
Motivation of this work is to propose a criterion that is free from those issues and
formulate the PBH abundance and the mass function based on the new criterion.

Before closing this section, we emphasize that we do not claim that our formulation is perfect and rigorous.
Yet, our formulation is an improvement from the previous studies in the sense that
it has no conceptual issue and at least captures a physically reasonable aspect that has not been taken into
account in the literature.

\section{Formulation of the PBH mass function}
\label{formulation}
In this section, we provide a new formulation of the PBH mass function.

\subsection{Variable to describe the PBH formation}
Qualitatively, physical mechanism of the PBH formation in the radiation dominated era is very simple, namely, the Jeans instability.
In other words, a PBH is formed when the self-gravity of an overdense region defeats the radiation pressure \cite{Hawking:1971ei}.
Yet, quantitative formulation of the criterion of the PBH formation, because of its nonlinear nature, is a non-trivial issue
and has been a topic of long-standing research.
A simple Newtonian analysis shows that the PBH is formed if the density contrast of the overdense region
evaluated at the time of the horizon reentry satisfies $\delta > \delta_{\rm th}=w$,
where $w=\frac{1}{3}$ is the equation of state parameter of radiation \cite{Carr:1975qj}.
A refined analysis gives 
$\delta_{\rm th} =\sin^2 [ \pi \sqrt{w}/(1+3w) ]$ \cite{Harada:2013epa}.
These formulae are derived under the assumption that a spherically symmetric overdensity is superposed
on top of the unperturbed Friedmann-Lema\^itre-Robertson-Walker (FLRW) universe and has a uniform overdensity
represented by $\delta$.
In reality, the density profile is not constant in general and numerical simulations is the only way to determine the
threshold, where it has been shown that the threshold value depends on the density profile \cite{Niemeyer:1999ak, Shibata:1999zs, Polnarev:2006aa, Nakama:2013ica, Musco:2018rwt} (see also \cite{Escriva:2019phb,Escriva:2019nsa}).
Furthermore, the primordial fluctuations are random field and those in real space consist of modes covering
a wide range of wavelength if the power spectrum is broad while most simulations deal with isolated overdensity.
However, this does not mean that all the fluctuation modes are relevant to the PBH formation.
Since the PBH is formed at the scale of the Hubble horizon, 
fluctuations which are super-Hubble at the time of the PBH formation are absorbed into redefinition of the background quantities and
should not affect the dynamics of the PBH formation.
Similarly, sub-Hubble fluctuations have already diminished when the larger-scale fluctuations reenter the Hubble
horizon and hence they should be also irrelevant.
The local quantities such as the density contrast on the comoving slice
are free from the super-Hubble modes and are a natural variable to satisfy the former condition \cite{Young:2014ana}.
Another local quantity useful to characterize the PBH formation is the compaction function ${\cal C} (r)$
\cite{Shibata:1999zs, Germani:2018jgr, Musco:2018rwt}.
This quantity is defined in terms of the mass excess of the overdense region.
Approximating the overdense region to be a part of the closed FLRW universe, 
the compaction function is roughly given by
${(L/r_{\rm curv})}^2$ on super-Hubble scales where $L$ is the proper size of the overdense region and $r_{\rm curv}$
is the curvature radius determined by the 3-d curvature. 
Since both $L$ and $r_{\rm curv}$ grow in proportional to the scale factor, the compaction function is conserved 
on super-Hubble scales.
To be definite, we use the density contrast on the comoving slice as the proper local quantity although
the our formulation can be applied to any other quantities such as the compaction function as long as it is as good as the density contrast.
The latter condition about the sub-Hubble modes is effectively solved by employing a window function (e.g. \cite{Young:2019osy}).
Thus, throughout this paper, we use the density contrast on the comoving slice smoothed by the window function
to compute the mass function.

Before closing this subsection, it may be useful to have a qualitative understanding of the smoothed density contrast.
At the linear order, the density contrast on the comoving slice is written by the
curvature perturbation ${\cal R}$ as
\be
\theta =-\frac{4}{9a^2 H^2} \triangle {\cal R}. \label{linear-t-R}
\ee
Here we use $\theta$ instead of $\delta$ to avoid unnecessary confusion with the Dirac's delta function which appears later.
Then, the smoothed density contrast corresponding to the horizon crossing $R=1/(aH)$ is given by
\be
\theta (R;{\vec x})=-\frac{4}{9} R^2 \int d^3y ~W (R; -{\vec x}+{\vec y}) \triangle {\cal R}({\vec y}),
\ee
where $R$ is the smoothing scale in the comoving unit and $W$ is the window function.
The power spectrum of the smoothed $\theta$ is given by
\be
P_{\theta (R)}(k)={\left( \frac{4}{9} \right)}^2 {(Rk)}^4 {\tilde W}^2 (R;k) P_{\cal R}(k),
\ee
where ${\tilde W}$ is the Fourier transform of $W$.
Due to the factor ${(Rk)}^4$, the power spectrum is suppressed on the super-Hubble scales $Rk \ll 1$.
Due to the window function, it is also suppressed on the sub-Hubble scales $Rk \gg 1$.
Thus, the smoothed density contrast is dominated by modes varying at the scales $\simeq R$ \footnote{
If one wants to use the compaction function as the basic variable instead of the density contrast,
we take $\theta (R;{\vec x})$ to be $C(R,{\vec x})$ which is a compaction function of the region having its center at ${\vec x}$ 
and its size $R$.}.
At the non-linear level, higher-order terms in ${\cal R}$ appear on the right-hand side of Eq.~(\ref{linear-t-R}), but
the qualitative features mentioned above still hold \footnote{Effects of the non-linear relation on the PBH abundance
were addressed in \cite{Kawasaki:2019mbl, Young:2019yug}.}. 

\subsection{About the previous studies}
\label{previous}
Having defined the variable to describe the PBH formation,
let us proceed to the discussion of the PBH mass function.
In order to be definite, we define the PBH mass function $f(M)$ such that $f(M)d\ln M$ is
the probability that a randomly chosen PBH has mass in $(M,M+d\ln M)$,
\be
\int f(M) d\ln M=1. \label{def-mass}
\ee
Thus, $f(M)$ represents relative fraction of PBHs in the logarithmic mass bin.
Before explaining our idea of formulating the mass function, let us briefly review the formalisms
used in the literature and their drawbacks.

In the literature, two approaches have been used to calculate the PBH mass function.
The first approach is the Press-Schechter formalism.
In this formalism, we introduce the fraction $\beta$ defined as \footnote{In some papers, an overall factor $2$ is multiplied on the right-hand side of 
Eq.~(\ref{PS-beta}) \cite{Kim:1996hr, Green:2004wb, Bugaev:2008gw}. In Eq.~(\ref{PS-beta}), we do not include the factor since its necessity is not clear.}
\be
\beta (R) =\int_{\rm \theta_{\rm th}}^\infty ~P(\theta_R) d\theta_R, \label{PS-beta}
\ee
where $P(\theta_R)$ is a probability density function of $\theta_R$  and $\theta_R$ is a shorthand notation of the smoothed density contrast $\theta (R;{\vec x})$ ($R$ is the smoothing scale),
and interpret it as a fraction of PBHs whose mass is larger than the one corresponding to 
$R$ \cite{Kim:1996hr, Green:2004wb, Bugaev:2008gw, Young:2014ana}.
Then the mass function is obtained by taking differential of $\beta$
with respect to $M$ (corresponding to the smoothing scale $R$)
\cite{Kim:1996hr}.
Although this prescription works well for the formation of the large scale structure,
its use is not justified for the PBH formation since PBHs at different masses form at different horizon-crossing times.
In Refs.~\cite{Carr:2009jm, Byrnes:2018clq, Byrnes:2018txb, Kawasaki:2019mbl, Young:2019yug, Wang:2019kaf}, 
the fraction $\beta$ itself was identified with the mass function (apart from a trivial evolution factor during the radiation dominated
era and the dark energy era).
However, validity of that identification, especially when the power spectrum is broad and PBHs are expected to
form over a wide mass range, 
is also not clear since $\beta$ is not defined as the quantity representing a differential fraction.
The root of this confusion lies at the lack of the criterion of the PBH formation which automatically enables us to 
derive the mass function given by Eq.~(\ref{def-mass}) without additional interpretation.

The second approach is the peak theory in which PBHs are supposed to form out of high-$\sigma$ peaks
of the primordial perturbations.
This approach has been used in Refs.~\cite{Green:2004wb, Young:2014ana, Yoo:2018esr, Germani:2018jgr,Kalaja:2019uju}. 
In Ref. \cite{Green:2004wb}, it was assumed that peaks of the perturbations smoothed at the scale $R$ that collapsed upon 
horizon reentry represents PBHs heavier than the ones corresponding to $R$.
Again, there is no clear justification for this prescription. 
In Ref.~\cite{Yoo:2018esr}, the PBH mass function is derived by relating the size and height of peaks with the PBH mass and
translating the size and height distribution of peaks to the distribution of the PBH mass. 
Yet, elimination of the sub-Hubble modes are not performed in their analysis and hence their method cannot be
directly applied to the broad power spectrum (see e.g. \cite{Young:2019osy, Young:2019yug}).
To summarize, to the best of our knowledge, there is no formulation of the PBH mass function in the literature
which are free from the conceptual issues mentioned above and automatically provides the differential mass function.

\subsection{A criterion of the PBH formation}
We have discussed limitations of the previous formulations to compute the PBH mass function.
Here we present a novel criterion of the PBH formation which is an addition of one extra condition to
the existing criterion.
We start from explaining a basic picture behind our idea.
It is often stated that the density fluctuations smoothed over the Hubble horizon collapse to PBHs if their 
amplitudes are greater than a threshold \cite{Carr:1975qj}. 
It is this point that we want to address here.
Consider an overdense region with the comoving size $R_*$ which collapses to a BH soon after the horizon reentry.
As we discussed previously, this happens if the amplitude of $\theta_{R_*}$ is greater than the threshold. 
By continuity, $\theta_R$ of the same region with slightly different value $R=R_*+dR$ would still exceed the threshold.
When $R$ deviates from $R_*$ by ${\cal O}(1)$, $\theta_R$ will no longer be larger than the threshold.
Although the range of $R$ for which $\theta_R$ is greater than the threshold is not infinitesimally narrow,
outcome of the horizon reentry of the overdense region is a formation of only a single PBH. 
In other words, a PBH is formed at the scale $R_*$ when $\theta_R$ is greater than the threshold in the small
interval $(R_-, R_+)$ where $R_- < R_* < R_+$.
Based on this observation, it is physically reasonable to assume that a PBH is formed at site where
not only spatial gradient of $\theta_R$ but also $\theta_{R,R}:=\partial\, \theta_R / \partial R$ vanish.
In order to translate this picture into equation, 
we identify $R_*$ with $R$ at which $\theta_{R,R}=0$.

To summarize, we adopt the following conditions as a criterion of the PBH formation.
A PBH is formed at point ${\vec x}$ satisfying the following conditions
\be
\theta_{R,a}=0,~~~\lambda_a (\theta_{R,ij}) <0,~~~\theta_R > \theta_{\rm th}, \label{PBH-criterion}
\ee
Here $a, b$ are indices for the spatial coordinates as well as $R$ that run from 1 to 4 and
$\lambda_a (\theta_{R,ij})$ are the eigenvalues of the $4\times 4$ matrix $\theta_{R,ab}$.
The second condition ensures that the peak is locally maximum.
The mass of the PBH is given by $M=m(R_*,\theta (R_*))$, where $R_*$ is a value of 
$R$ at which the above conditions are satisfied. 
We include the dependence of the PBH mass $m$ on $\theta$ to account for the critical collapse.

\subsection{PBH mass function}
\begin{figure}[tbp]
  \begin{center}
   \includegraphics[width=100mm]{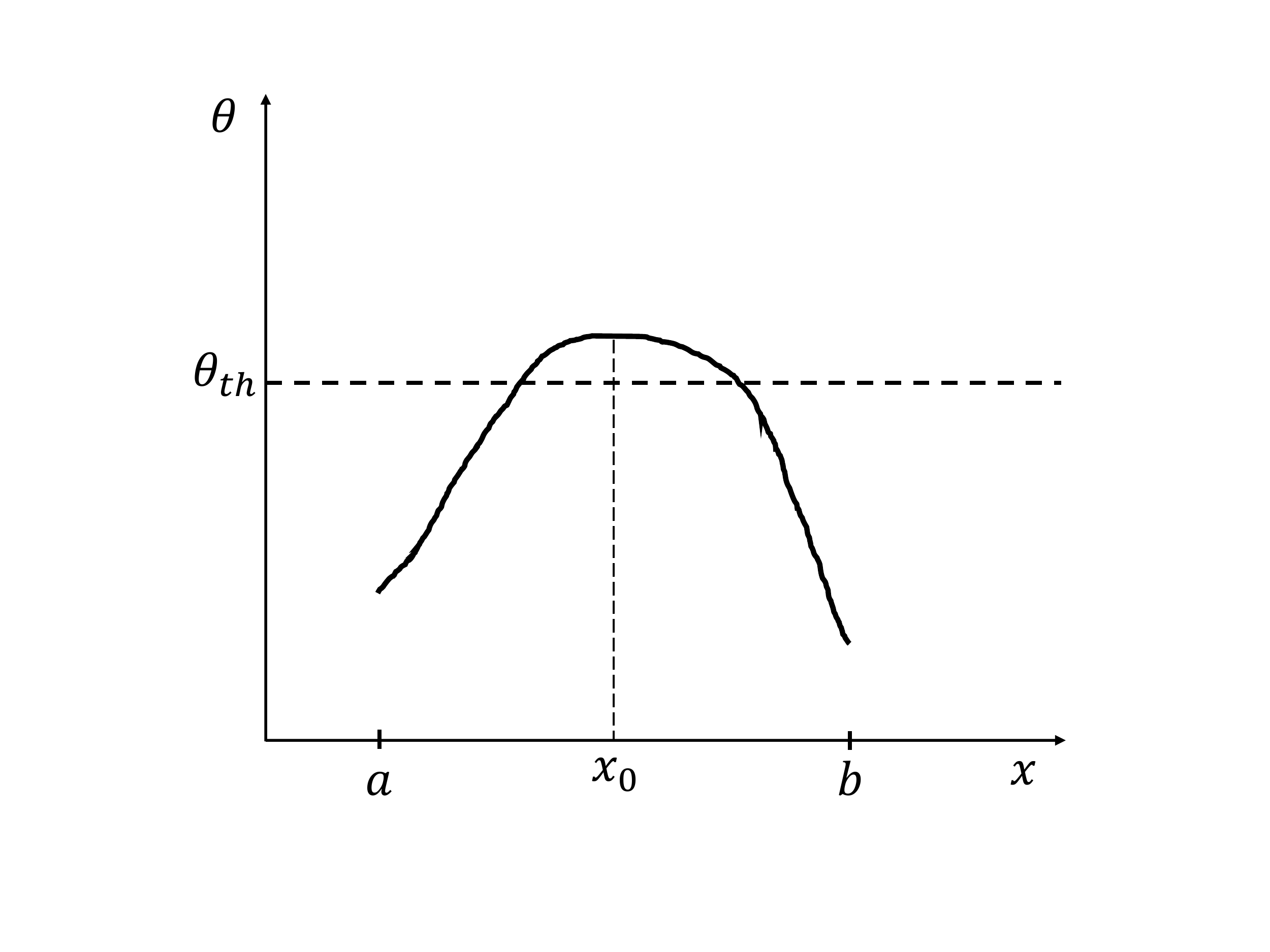}
  \end{center}
  \caption{A PBH is formed at $x=x_0$ where $\theta_{R,x}=0$ and $\theta_{R,xx}<0$. }
  \label{toy}
\end{figure}

Having defined the criterion of the PBH formation by Eq.~(\ref{PBH-criterion}),
we can now formulate the PBH mass function.
In order to clarify the technical points, let us start with a very simple model in which
the space dimension is one, perturbations are nearly monochromatic (at $R_*$), and the effect of the critical phenomena is ignored ($m=m(R_*)$).
In this case, we may ignore the $R$-dependence, and in stead of Eq.~(\ref{PBH-criterion}), 
we can assume that a PBH is formed at a point satisfying
\be
\theta_{R,x}=0,~~~~\theta_{R,xx}<0,~~~~\theta_R > \theta_{\rm th}.
\ee
Let us consider a configuration $\theta_R (x)$ shown in Fig.~\ref{toy}.
In this case, in an interval $a\le x \le b$, a PBH is formed at $x=x_0$ where $\theta_{R,x}=0$.
Using the Dirac's delta-function and the Heaviside function $\Theta$, the number of PBHs (in the present example, it is one) is expressed as
\be
1=\int_a^b \delta (\theta_{R,x}) \Theta (\theta_R-\theta_{\rm th})|\theta_{R,xx}| \Theta (-\theta_{R,xx}) dx. \label{one}
\ee
Here $|\theta_{R,xx}|$ needs to be inserted to compensate the factor coming from the argument of
the delta-function.
Thus, for the present configuration, the PBH number density is given by
\be
\delta (\theta_{R,x}) \Theta (\theta_R-\theta_{\rm th}) |\theta_{R,xx}| \Theta (-\theta_{R,xx}).
\ee
In reality, $\theta$ is a random variable and the PBH number density must be given by the ensemble average
of the above quantity.
Hence, the comoving number density of PBH is given by
\be
n_{\rm PBH}=\langle \delta (\theta_{R,x}) \Theta (\theta_R-\theta_{\rm th}) |\theta_{R,xx}| \Theta (-\theta_{R,xx}) \rangle \equiv
\int [d\theta] \delta (\theta_{R,x}) \Theta (\theta_R-\theta_{\rm th}) |\theta_{R,xx}| \Theta (-\theta_{R,xx}) P[\theta],
\ee
where the normalization of the functional integral is defined by
\be
\int [d\theta] P[\theta]=1.
\ee
In this case, the PBH mass function defined by Eq.~(\ref{def-mass}) is given by
\be
f(M)=m_* \delta (M-m_*),
\ee
where $m_* \equiv m(R_*)$.

As a second simple example, let us recover the $R$-dependence and assume that a PBH is formed at a point satisfying
\be
\theta_{R,x}=0,~~~~\theta_{R,R}=0,~~~~\theta_{R,xx}+\theta_{R,RR}<0,~~~~\theta_{R,xx}\theta_{R,RR}-\theta_{R,xR}^2>0~~~~~\theta_R > \theta_{\rm th}.
\ee
In the same way as the above example, the number density of PBH is given by
\be
n_{\rm PBH}=\int dR dM~ \langle 
J \delta (\theta_{R,x}) \delta (\theta_{R,R}) \delta (M-m(R,\theta_R)) \Theta (\theta_R-\theta_{\rm th}) 
\Theta (-\theta_{R,xx}-\theta_{R,RR})
\Theta (\theta_{R,xx}\theta_{R,RR}-\theta_{R,xR}^2) \rangle,
\ee
where $J$ defined by
\be
J=\begin{vmatrix}
\theta_{R,xx} & \theta_{R,xR} \\
\theta_{R,Rx} & \theta_{R,RR} \\
\end{vmatrix} .
\ee
accounts for the Jacobian needed to compensate the factor coming from the argument of the delta-functions for $\theta_{R,x}$
and $\theta_{R,R}$.
Thus, the PBH mass function is written as
\be
f(M)=\frac{M}{n_{\rm PBH}}\int dR~ \langle 
J \delta (\theta_{R,x}) \delta (\theta_{R,R}) \delta (M-m(R,\theta_R)) \Theta (\theta_R-\theta_{\rm th}) 
\Theta (-\theta_{R,xx}-\theta_{R,RR}) \Theta (\theta_{R,xx}\theta_{R,RR}-\theta_{R,xR}^2)\rangle. \label{1d-mass}
\ee

After these examples, it is immediate to generalize the mass function (\ref{1d-mass}) to that in the three dimensional space.
What we need to do is simply to substitute $J$ given by
\be
J=\begin{vmatrix}
\theta_{R,xx} & \theta_{R,xy} & \theta_{R,xz} & \theta_{R,xR} \\
\theta_{R,yx} & \theta_{R,yy} & \theta_{R,yz} & \theta_{R,yR} \\
\theta_{R,zx} & \theta_{R,zy} & \theta_{R,zz} & \theta_{R,zR} \\
\theta_{R,Rx} & \theta_{R,Ry} & \theta_{R,Rz} & \theta_{R,RR} \\
\end{vmatrix}.
\ee
Thus, the PBH mass function is formally given by
\be
f(M)=\frac{M}{n_{\rm PBH}}\int dR~ \langle 
J \delta (M-m(R,\theta_R)) \Theta (\theta_R-\theta_{\rm th}) 
\prod_{a=1}^4 \delta (\theta_{R,a})  \Theta (-\lambda_a) \rangle.  \label{3d-mass}
\ee
Eq.~(\ref{3d-mass}) is the main result of this paper.
There are a few remarks worth mentioning at this stage.
First, the right-hand side is an ensemble average of a quantity consisting of $\theta$ 
evaluated at a fixed point ${\vec x}$.
Thus, in nature Eq.~(\ref{3d-mass}) is dependent on ${\vec x}$.
Yet, when the perturbation $\theta$ respects the translation symmetry, which is satisfied in most situations,
Eq.~(\ref{3d-mass})  becomes independent of ${\vec x}$.
In such a case, we can choose any point ${\vec x}$ to evaluate the mass function.
Secondly, if the effect of the critical phenomenon can be ignored \footnote{The mass function
due to the critical phenomena for the monochromatic power spectrum of the Gaussian primordial perturbations
was derived in \cite{Niemeyer:1997mt} and later extended to the non-Gaussian case in \cite{Yokoyama:1998xd}.},
which may be a good approximation if the mass function in the mass range of our interest represents PBHs that
have nearly horizon mass at their formation time, we can easily perform the integration over $R$
by dropping the dependence on $\theta_R$.
Writing the PBH mass as $M=M(R)$, Eq.~(\ref{3d-mass}) after the integration becomes
\be
f(M)=\frac{M}{n_{\rm PBH}} \frac{dR}{dM} \langle 
J  \Theta (\theta_R-\theta_{\rm th}) \prod_{a=1}^4 \delta (\theta_{R,a}) \Theta (-\lambda_a) \rangle. 
\label{eq:non-critical-mass}
\ee
Thirdly, Eq.~(\ref{3d-mass}) can be applied to any type of primordial perturbations.
All the information about the statistical properties of $\theta$ is encoded in $P[\theta]$ and
the mass function is given by the functional integration with the probability weight $P[\theta]$.
Thus, for a given $P[\theta]$, we can in principle compute the mass function
by performing the functional integration of Eq.~(\ref{3d-mass}) although the feasibility of the 
practical implementation will depend on concrete form of $P[\theta]$.
Finally, we do not claim that Eq.~(\ref{3d-mass}) provides perfectly correct PBH mass function
for given $P[\theta]$.
In deriving the result, we have made several approximations which may invalidate Eq.~(\ref{3d-mass})
if ones wants to obtain more precise mass function.
First, our formula contains the window function and the mass function will change as we change the window function
(see  \cite{Young:2019osy} and \cite{Ando:2018qdb} which discuss about the sensitivity of the result to the choice
of the window functions.).
Unfortunately, there is no definite principle to choose the best window function.
It is not even clear if there exists the universal window function valid for any type of $P[\theta]$ and
it is possible that the prescription of using the smoothed perturbations has limitation and never
gives the exactly correct mass function.
Even if we accept the smoothing prescription, 
the PBH formation is not simply judged by a single universal value $\theta_{\rm th}$ as the threshold
depends on the perturbation profile.
Another issue is that the point where $\theta_{R,R}=0$ will not be precisely corresponding to the scale of the PBH formation. 
Despite of these drawbacks, our criterion of the PBH formation (\ref{PBH-criterion}) and the
formula of the mass function (\ref{3d-mass}) go beyond the previous studies in the sense that
our criterion captures a basic physical picture of the PBH formation, is well-defined mathematically,
and derives the mass function automatically while the previous results have difficulty of deriving 
the mass function at the conceptual level as explained in \ref{previous}.

\section{Application to the Gaussian perturbations}
\label{gaussian}
As a demonstration of the new formalism, we compute the PBH mass function for a simple case
where the space dimension is 1, probability weight of $\theta$ is Gaussian, and the critical collapse is ignored.
With these assumptions, the mass function is given by (see Eq.~(\ref{eq:non-critical-mass}))
\be
f(M)=\frac{M}{n_{\rm PBH} m'(R)}\langle J \delta (\theta_{R,x}) \delta (\theta_{R,R}) \Theta (\theta_R-\theta_{\rm th}) 
\Theta (-\theta_{R,xx}-\theta_{R,RR}) \Theta (\theta_{R,xx} \theta_{R,RR}-\theta_{R,xR}^2) \rangle.
\ee
In the last expression, it should be understood that $R$ on the right-hand side is determined by a relation $M=m(R)$.
In order not to terminate the radiation dominated epoch before the time of the matter-radiation equality,
PBHs must be only a tiny fraction of the energy density of the Universe at the time when they are produced.
This means the threshold $\theta_{\rm th}$ is much larger than the mean amplitude of $\theta_R$.
In other words, PBHs are formed only out of extremely high-$\sigma$ fluctuations.
In such a situation, site at which $\theta_{R,x}=\theta_{R,R}=0$ is satisfied would satisfy
$\theta_{R,xx}+\theta_{R,RR}<0$ and $J=$ $\theta_{R,xx} \theta_{R,RR}-\theta_{R,xR}^2 >0$ with probability close to unity.
Then, to a good approximation, the above mass function is written as
\be
f(M)=\frac{M}{n_{\rm PBH} m'(R)} I(M), ~~~~~
I(M)= \langle  (\theta_{R,xx} \theta_{R,RR}-\theta_{R,xR}^2) 
\delta (\theta_{R,x}) \delta (\theta_{R,R}) \Theta (\theta_R-\theta_{\rm th}) \rangle.
\ee
Now, the problem has been reduced to evaluation of $I(M)$.

Since both the step function and the $\delta$-function appearing in $I(M)$ are not convenient quantities 
to compute the correlation functions,
let us replace them by using the following formulae
\be
\delta (\theta_{R,x})=\int \frac{d\eta_1}{2\pi} e^{i\eta_1 \theta_{R,x}},~~~
\delta (\theta_{R,R})=\int \frac{d\eta_2}{2\pi} e^{i\eta_2 \theta_{R,R}},~~~
\Theta (\theta_R-\theta_{\rm th})= \int_{\theta_{\rm th}}^\infty d\alpha \int \frac{d\eta_3}{2\pi} e^{i\eta_3 (\theta_R-\alpha)}.
\ee
The result is given by
\be
I(M)=\int \frac{d\eta_1}{2\pi} \int \frac{d\eta_2}{2\pi} \int \frac{d\eta_3}{2\pi} \int_{\theta_{\rm th}}^\infty d\alpha e^{-i\eta_3 \alpha}
\langle  (\theta_{R,xx} \theta_{R,RR}-\theta_{R,xR}^2) e^{i\eta_1 \theta_{R,x}+i\eta_2 \theta_{R,R}+i\eta_3 \theta_R} \rangle. \label{eq-I-1}
\ee
The integrand in the above expression is given by correlations of a particular combination of the six Gaussian variables 
$\{ \theta_R, \theta_{R,x}, \theta_{R,R}, \theta_{R,xx}, \theta_{R,RR}, \theta_{R,xR} \}$.
These variables are correlated with each other.
For instance, in terms of the power spectrum of $\theta$ and the window function $W$,
off-diagonal components of the covariance matrix
can be written as
\be
\langle \theta_R \theta_{R,xx} \rangle =-\int \frac{dk}{2\pi} {\tilde W}^2(R;k) k^2P_\theta (k),~~~
\langle \theta_R \theta_{R,R} \rangle=\int \frac{dk}{2\pi} {\tilde W} (R;k) {\tilde W}_R (R;k) P_\theta (k),
\ee
with $\tilde{W}_R:=d\tilde{W}/dR$ and so on, which do not vanish in general.
Since any correlator of Gaussian variables is determined completely in terms of their covariance,
$I(M)$ should also be expressed in terms of the covariance of the six variables.
For any Gaussian variables $x_i$ with zero mean and covariance $\langle x_i x_j \rangle=P_{ij}$,
we have
\be
\langle x_i x_j e^{iv_k x_k} \rangle = (P_{ij}-v_k P_{ki} v_\ell P_{\ell j} ) 
\exp \left( -\frac{1}{2} P_{k\ell} v_k v_\ell \right).
\ee
Applying this formula to Eq.~(\ref{eq-I-1}), we obtain
\be
I(M)=\int_{\theta_{\rm th}}^\infty d\alpha \int \frac{d^3 \eta}{{(2\pi)}^3}
(A+B_{ij} \eta_i \eta_j) \exp \left( -\frac{1}{2} M_{ij} \eta_i \eta_j+i v_i \eta_i \right),
\ee
where $A, B_{ij}, M_{ij}, v_i$ are given by
\begin{align}
A&=\langle \theta_{R,xx} \theta_{R,RR}\rangle-\langle \theta_{R,xR}^2 \rangle, \\
B_{ij}&= \begin{cases}
    {\langle \theta_{R,x} \theta_{R,xR} \rangle}^2 & (i=j=1) \\
    -\langle \theta_{R,R} \theta_{R,xx} \rangle \langle \theta_{R,R} \theta_{R,RR} \rangle & (i=j=2) \\
    - \langle \theta_R \theta_{R,xx} \rangle \langle \theta_R \theta_{R,RR} \rangle & (i=j=3)\\
     -\frac{1}{2} \left( \langle \theta_R \theta_{R,xx} \rangle
\langle \theta_{R,R} \theta_{R,RR} \rangle + \langle \theta_{R,R} \theta_{R,xx} \rangle \langle \theta_R \theta_{R,RR} \rangle \right) & (i=2,j=3~{\rm or}~i=3,j=2)\\
    0 & ({\rm otherwise})
  \end{cases} \\
M_{ij}&=\begin{pmatrix}
\langle \theta_{R,x}^2 \rangle & 0 & 0\\
0 & \langle \theta_{R,R}^2 \rangle & \langle \theta_R \theta_{R,R} \rangle \\
0 & \langle \theta_R \theta_{R,R} \rangle & \langle \theta^2_R \rangle \\
\end{pmatrix} \\
v_i&=(0,0,-\alpha).
\end{align}

Integration over $\eta_i$ yields
\be
I(M)={\left( \det \left( 2\pi M \right) \right)}^{-\frac{1}{2}}
\int_{\theta_{\rm th}}^\infty d\alpha ~(A+B_{ij} N_{ij}-B_{ij} N_{i3}N_{j3} \alpha^2 )
\exp \left( -\frac{1}{2} N_{33} \alpha^2 \right),
\ee
where $N_{ij}$ is inverse matrix of $M_{ij}$.
Finally, the integration over $\alpha$ can be expressed in terms of the complementary error function as
\begin{align}
I(M)=&{\left( \det \left( 2\pi M \right) \right)}^{-\frac{1}{2}} \sqrt{\frac{2}{N_{33}}}
\bigg[ \left( A+B_{ij}N_{ij}-\frac{B_{ij} N_{i3} N_{j3}}{N_{33}} \right) \frac{\sqrt{\pi}}{2}
{\rm erfc} \left( \sqrt{\frac{N_{33}}{2}} \theta_{\rm th} \right) \nonumber \\
&-\frac{B_{ij} N_{i3}N_{j3}}{\sqrt{2N_{33}}} \theta_{\rm th} \exp \left( -\frac{N_{33}}{2} \theta_{\rm th}^2 \right) \bigg],
\end{align}
where 
\be
{\rm erfc}(x)=\frac{2}{\sqrt{\pi}} \int_x^\infty e^{-t^2}dt.
\ee

\section{Conclusion}
Formulation and computation of the PBH mass function from the sourcing primordial density perturbations 
have been addressed in the literature. 
As we have discussed in depth in Sec.~\ref{formulation},
the formulated PBH mass function given in the literature have conceptual issues which are ambiguities 
that arise in translating the formation probability into the PBH mass function which is differential quantity.
Although the formulations in the previous studies would be practically acceptable when the power spectrum
of the perturbations are sharply localized around a particular scale, 
the ambiguities become more prominent when one considers the broad power spectrum.
This issue stems from the lack of the physically understandable and quantitative criterion of the PBH formation
which automatically leads to the PBH mass function without bringing up the artificial interpretations
adopted in the literature.

In this paper, we proposed a new criterion of the PBH formation which is an addition of one extra condition
to the existing criterion.
By doing this, we formulated the PBH mass function, which is given by Eq.~(\ref{3d-mass}), 
without introducing any ambiguities.
Eq.~(\ref{3d-mass}) is a formal expression defined as an ensemble average of the integrand
weighted by the functional probability density $P[\theta]$. 
Once $P[\theta]$ is given, the PBH mass function can be in principle determined by performing
the functional integral defined by Eq.~(\ref{3d-mass}).
Practically, analytic evaluation is feasible only for very limited cases such as Gaussian perturbation.
In Sec.~\ref{gaussian}, as a demonstration of our formulation,
we computed the PBH mass function analytically for the case where the perturbations are Gaussian
and the space is one dimension.

Finally but not least, it would be possible to apply our criterion to formulate other statistical quantities describing PBHs.
For instance, spin distribution of PBHs (see \cite{Chiba:2017rvs, Mirbabayi:2019uph, DeLuca:2019buf}) will be obtained 
by replacing the $\delta$-function
for $M-m(R,\theta)$ in Eq.~(\ref{3d-mass}) with $J-j(R,\theta)$ where $j(R,\theta)$ is the angular momentum
of the overdense region parameterized by $R$ and $\theta$ and by
promoting the threshold $\theta_{\rm th}$ to spin-dependent function \cite{He:2019cdb}.
Two-point correlation functions of PBHs is another important quantity since it describes clustering of PBHs \cite{Suyama:2019cst}.
This may be also obtained by applying the criterion at two different points simultaneously.

\section*{Acknowledgement}
TS is supported by JSPS Grant-in-Aid for Young Scientists (B) No.15K17632,  by the MEXT Grant-in-Aid for Scientific Research on In-novative Areas No.15H05888, No.17H06359, No.18H04338, and No.19K03864. SY is supported
by MEXT KAKENHI Grant Numbers 15H05888 and 18H04356.

\bibliographystyle{utphys}
\bibliography{draft}

\end{document}